\documentclass[aps,prc,groupedaddress,showpacs,manuscript]{revtex4}

\usepackage{graphicx}

\begin{document}

\title{Probing the density dependence of the symmetry potential in intermediate energy heavy ion collisions}

\author {Qingfeng Li$\, ^{1}$\footnote{Fellow of the Alexander von Humboldt Foundation.}
\email[]{Qi.Li@fias.uni-frankfurt.de}, Zhuxia Li$\,^{1,2}$
\email[]{lizwux@iris.ciae.ac.cn}, Sven Soff$\,^{3}$,
\\
Raj K. Gupta$\,^{1,4}$\footnote{DFG Mercator Guest Professor.}, Marcus Bleicher$\,^{3}$, and
Horst St\"{o}cker$\,^{1,3}$}
\address{
1) Frankfurt Institute for Advanced Studies (FIAS), Johann Wolfgang Goethe-Universit\"{a}t, Max-von-Laue-Str.\ 1, D-60438 Frankfurt am Main, Germany\\
2) China Institute of Atomic Energy, P.O. Box 275 (18),
Beijing 102413, P.R. China\\
3) Institut f\"{u}r Theoretische Physik, Johann Wolfgang Goethe-Universit\"{a}t, Max-von-Laue-Str.\ 1, D-60438 Frankfurt am Main, Germany\\
4) Department of Physics, Panjab University, Chandigarh -- 160014, India.
 }


\begin{abstract}
Based on the ultrarelativistic quantum molecular dynamics (UrQMD)
model, the effects of the density-dependent symmetry potential
for baryons and of the Coulomb potential
for produced mesons are investigated for neutron-rich heavy ion collisions at intermediate energies. The
calculated results of the $\Delta^-/\Delta^{++}$ and $\pi ^{-}/\pi
^{+}$ production ratios show a clear beam-energy dependence on the
density-dependent symmetry potential, which is stronger for
the $\pi ^{-}/\pi ^{+}$ ratio close to the pion
production threshold. The Coulomb potential of the mesons changes the
transverse momentum distribution of the $\pi ^{-}/\pi ^{+}$ ratio
significantly, though it alters only slightly the $\pi^-$ and
$\pi^+$ total yields. The $\pi^-$ yields, especially at
midrapidity or at low transverse momenta and the
$\pi^-/\pi^+$ ratios at low transverse momenta, are shown to be
sensitive probes of the density-dependent symmetry potential in
dense nuclear matter. The effect of the density-dependent
symmetry potential on the production of both, K$^0$ and K$^+$ mesons, is also investigated.
\end{abstract}


\maketitle

\section{Introduction}
The study of isospin effects in nuclear matter or finite nuclei is
a well established topic in nuclear physics, as well as in nuclear
astrophysics \cite{Pra97,Lat01}. On the mean field level, this
means the contribution of the symmetry energy of particles to the
total energy (see, e.g., the recent Refs.
\cite{Cha97,LiB02,LiB05,Gai04,Gai042,Riz04,LiQ01,LiQ05}), and on
the two-body collision level, detailed experimental data on the
free proton-proton and neutron-proton scattering cross sections
have been obtained \cite{PDG96} as a function of energy ranging from the Coulomb barrier to ultra high energy interactions. At low energies (up to several hundreds of
MeV/nucleon), large differences are visible between the
proton-proton and the neutron-proton cross sections. In
heavy ion collisions (HICs), it is also necessary to study the effect of density on the
isospin-dependent nucleon-nucleon cross sections.

Recently, we have explored the density- and temperature-dependence
of the nucleon-nucleon elastic scattering cross sections in
collisions between neutron-rich nuclei at intermediate energies
\cite{LiQ04}. We found a density dependence of the elastic
scattering cross sections. This work is based on the theory of
quantum hydrodynamics (QHD), in which the interaction between
nucleons is described by the exchange of $\sigma$, $\omega$,
$\rho$, and $\delta$ mesons. Since, in the QHD theory, both the
mean field and the collision term originate from the same
Lagrangian density, the density-dependence of the
symmetry energy is our next subject to investigate. Currently the exploration of the isospin dependence of the nuclear interaction (
especially in exotic and very isospin-asymmetric systems) has gained high interest \cite{Akm97,Hei00,LiB02,LiB05,Gai04,Gai042,LiQ05}. In particular,
the density dependence of the symmetry energy in {\it dense}
nuclear matter has to be studied in detail, because it suffers from large
uncertainties in the predictions of various theoretical models
\cite{Bro00,Mar01}. Based on the QHD theory, with a similar
effective Lagrangian density \cite{Hof01,LiuB02,Gai04}, the main
uncertainty of the density dependence of the symmetry energy results
from the uncertainties of the density-dependent coupling strengths
of various meson-baryon interactions, especially that of $\rho-N$
and $\delta-N$ interactions \cite{Gai04}. Also, in forthcoming
experiments at the Rare Isotope Accelerator (RIA) laboratory (USA)
and at the new international accelerator facility FAIR (Facility for Antiproton and Ion Research) at the Gesellschaft f\"{u}r Schwerionenforschung (GSI, Germany), more neutron-rich beams are
planned to be adopted, and hence call for further theoretical
studies on isospin-dependent HICs in the intermediate energy region.

In order to obtain detailed information on the density dependence
of the symmetry energy in dense nuclear matter, quite a few sensitive probes were brought up for experiments in recent years:
e.g., the $\pi^-/\pi^+$ yield ratio, the pion flow, the
transverse momentum distribution of $\pi^-/\pi^+$ ratios, the
neutron-proton differential flow \cite{LiB02}, and the isospin
equilibration and stopping \cite{Bass98,Rami00,Gai042}. More recently, we have
also investigated \cite{LiQ05} the sensitivity of the
$\Sigma^-/\Sigma^+$ ratio on the density-dependent symmetry
potentials. Further investigations seem to be worthwhile to shed light on
the rather unclear isospin-independent and -dependent mean-field potentials
of the hyperons. Since the
contribution of the symmetry energy to the whole dynamics of HICs is rather small at
SchwerIonen Synchrotron (SIS) energies, it is
all the more difficult to explore the density dependence of the
symmetry energy based on the current experimental situation.
Hence, in theory, more sensitive probes should be searched for.

Also, such a detailed analysis is quite necessary and important
because of the following arising questions: Is the effect of the density-dependent symmetry energy on the $\pi^{-}/\pi
^{+}$ ratio influenced by the beam energy? In
which momentum, rapidity, or phase-space, region are the pion
yields more suitable to probe the density-dependent symmetry
energy? And, how does the Coulomb potential of mesons affect the
emission of pions?  Gaitanos {\it et al.} \cite{Gai04,Gai042}
found that for beam energies higher than about $2A$ GeV, the
sensitivity of the $\pi^{-}/\pi ^{+}$ ratio to the form of the
symmetry energy at high densities is strongly reduced. Furthermore,
they found that the $\pi^-/\pi^+$ ratio at high transverse momenta
$p_t$ could offer sound information on the density dependence of
the symmetry energy. B.-A. Li {\it et al.} \cite{LiB05} revisited
the $\pi^-/\pi^+$ ratio by using an up-dated, new version of the
isospin-dependent Boltzmann-Uehling-Uhlenbeck (IBUU04) model, in
which an isospin- and momentum-dependent single nucleon potential
was adopted. They found that the sensitivity of the $\pi^-/\pi^+$
ratio on the density-dependence of the symmetry energy becomes obvious
after considering the momentum-dependent single nucleon potential.
They also noticed the effect of the Coulomb potential on the $\pi$
production, especially on the transverse momentum spectra. Their
work implied that it is sufficient to measure accurately the
low-energy (or the low-transverse-momentum $p_t$) pions, instead
of the whole spectrum. In this paper, we address these aspects of
the problem on the basis of the ultrarelativistic quantum
molecular dynamics (UrQMD) model.

In this work, we also investigate the $K^0/K^+$
production ratio as a probe of the density-dependent symmetry
potential in dense nuclear matter. Based on the constituent quark picture the $K^0$ meson contains one $d$-quark
while the $K^+$ contains one $u$-quark, as well as one
$\overline{s}$-quark in each meson. This is closely related
to the isospin asymmetry of the nuclear medium at SIS energies.
Furthermore, experimentally, the effects of system parameters, such as
beam energy or impact parameter, on the sensitive probes
also have to be kept in mind. Theoretically, the
uncertainty of the isospin-independent nuclear
equation of state (EoS), as well as the contribution of the
Coulomb potential of mesons, might alter the calculated
results of these probes, hence, they could even modify the conclusions and should be checked carefully.

In the UrQMD model (version 1.3)
\cite{Bass98,Bleicher99,Web03,Bra04}, adopted here, we find that
most of the calculations can simultaneously reproduce many
experimental measurements, which offers a good starting point for
studying the isospin effects at SIS energies. In this work, a
'hard' and a 'soft' Skyrme-type EoS, without momentum
dependence, are adopted for central ($b=0-2\ {\rm fm}$) or
near-peripheral ($b=5-8\ {\rm fm}$) $^{132}Sn+^{132}Sn$ reactions at
beam energies $E_{\rm b}=0.5A$ and $1.5A$ GeV. We consider phenomenological density-dependent symmetry potentials, which
will be discussed in more detail in the next section. The Coulomb
potentials of the mesons are switched on or off in order to
analyze more conveniently the contributions of the various
symmetry potentials to the dynamical evolution of the hadrons in
the nuclear medium.

The paper is arranged as follows: In Section II, we clarify the inclusion of the isospin-dependent part of the mean field
in the UrQMD transport model. In Section III, some basic isospin
effects on the dynamics of the baryons (here the nucleons and
$\Delta(1232)$) in SIS energy HICs are shown. In Section IV,
firstly the effects of the isospin-independent EoS, the beam
energy, and the impact parameter on the emitted pion yields and
the $\pi^-/\pi^+$ production ratios are discussed. Then, the
various phase-space distributions of pion yields and the
$\pi^-/\pi^+$ ratios, with the different density dependences of the
symmetry potentials, are investigated. At the end of this section,
the sensitivity of the $K^0/K^+$ ratio to the density dependence of
the symmetry potential is discussed. Finally, the
conclusions and outlook are given in Section V.

\section{The treatment of the potential update of hadrons in the UrQMD model}
The initialization of neutrons and protons, the
corresponding Pauli blocking, the Coulomb potential of baryons, and the isospin dependence of the nucleon-nucleon cross
sections have been introduced explicitly in the standard UrQMD
model. In order to study the isospin effects in intermediate
energy HICs, it is required to further introduce the
symmetry potential of baryons.

The isospin-dependent EoS for asymmetric nuclear matter can be
expressed as (see, {\it e.g.}, \cite{Bom91})
\begin{equation}
e(u,\delta)=\frac{\epsilon(u,\delta)}{\rho}=e_0(u)+e_{\rm sym}(u)\delta^2
,
\label{ieos}
\end{equation}
where $u=\rho/\rho_0$ is the reduced nuclear density and
$\delta=(\rho_n-\rho_p)/\rho$ is the isospin asymmetry in terms of
neutron ($\rho_n$) and proton ($\rho_p$) densities. The $e_0(u)$
term is the isospin-independent part, which includes the Skyrme
and the Yukawa potentials in the UrQMD model  \cite{Bass98}. The
Yukawa parameter is related to the Skyrme parameters since in infinite
nuclear matter the contribution of the Yukawa potential to the total
energy acts like a two-body Skyrme contribution. For
comparison, both a soft ($K=200$ MeV) and a
hard ($K=300$ MeV, the default parametrization in this work) EoS
are adopted in this work. $e_{\rm sym}$ is the symmetry energy per nucleon, in which the kinetic ($v_{\rm sym}^{\rm kin}$) and potential ($v_{\rm sym}^{\rm pot}$) contributions are included.
The average symmetry potential energy can be expressed as
\begin{equation}
v_{\rm sym}^{\rm pot}=e_{\rm a} F(u) \label{vsym},
\end{equation}
where $e_{\rm a}$ is the symmetry-potential
strength and $F(u)$ is the density-dependent part. If the Fermi-gas
model is adopted, the symmetry-potential strength $e_a$ is related to
the symmetry energy at normal density, $S_0$, by
\begin{equation}
S_0\simeq e_{\rm a}+\frac{\epsilon_F}{3}
. \label{s0ea}
\end{equation}
Here, $\epsilon_F$ is the Fermi kinetic energy at normal nuclear density.

The relativistic mean field calculations \cite{Gai04} show a small
variation of the symmetry kinetic energy extracted from various
models, and the uncertainty of the symmetry energy with nuclear
density results mainly from the symmetry potential energy.
Concerning the symmetry potential energy, both the symmetry energy coefficient $S_0$ and the
density dependence of the symmetry energy are quite uncertain. For $S_0$, deduced from
the isovector GDR in $^{208}Pb$ and from the available data of
differences between neutron and proton radii for $^{208}Pb$ and
several $Sn$ isotopes, a rather small range of values is $32\ {\rm MeV} \leq
S_0 \leq 36\ {\rm MeV}$ \cite{Vre03}. More recently, a new value of
$S_0\simeq31$ MeV was obtained, based on the consistent folding
analysis of the $p(^{6}He,^{6}He)p$ elastic scattering and
$p(^{6}He,^{6}Li^*)n$ charge exchange reaction data measured at
$E_{\rm Lab}=41.6$ MeV \cite{Kho05}, while $S_0=34$ MeV was obtained
in a calculation made within the relativistic  Brueckner framework,
using the {\rm Bonn A} potential \cite{Dal04}. Thus, it is necessary to
further investigate the uncertainty of the $S_0$ value \cite{LiQ04a}.
In this paper, however, the value $S_0=34$ MeV is adopted, since
we endeavor to investigate the density dependence of the
symmetry potential energy.

\begin{figure}
\includegraphics[angle=0,width=0.8\textwidth]{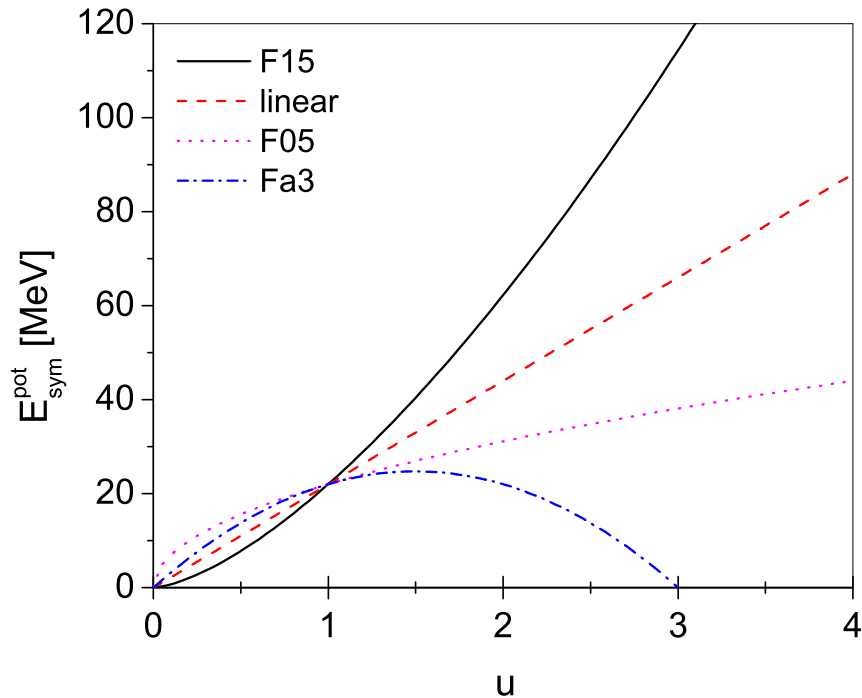}
\caption{The symmetry potential energy of nucleons as a function
of the reduced nuclear density $u$, using various density dependences (see text).}
\label{fig1}
\end{figure}

In order to mimic the strong density dependence of the symmetry
potential at high densities, we adopt the form of $F(u)$,
used in \cite{LiB02}, as
\begin{equation}
F(u)=\left\{
\begin{array}{l}
F_1=u^\gamma \hspace{1cm}  \gamma>0 \\
F_2=u\cdot\frac{a-u}{a-1} \hspace{1cm}a>1
\end{array}
\right. .\label{fu}
\end{equation}
Here, $\gamma$ is the strength of the density dependence
of the symmetry potential. We choose $\gamma=0.5$
and $1.5$, denoted as the symmetry potentials F05
and F15, respectively. $a$ (in F$_2$) is the reduced critical density; for
$u>a$, the symmetry potential energy is negative. We adopt
$a=3$ in this paper and the respective symmetry potential is named
as Fa3. The symmetry potentials F05, F15, and Fa3 are shown in
Fig.\ \ref{fig1} as a function of the reduced nuclear density $u$,
compared with a linear density-dependent symmetry potential.

From Fig.\ \ref{fig1}, it is apparent that for $u<1$,
F05$>$Fa3$>$F15, whereas for $u>1$, F15$>$F05$>$Fa3. F05 is larger
than Fa3 for all densities. This means that for
neutron-rich HICs at intermediate energies, less neutrons are pushed into the low density
region ($u<1$) as well as into the high density region ($u>1$) for density dependence
Fa3, as compared to F05. In other
words, the symmetry-potential parametrization
F05 is always stiffer than Fa3 (at both subnormal and higher densities), except at normal nuclear density.

Besides nucleons (N), the resonances N$^*$(1440), $\Delta$(1232),
and the hyperons $\Lambda$ and $\Sigma$ should be considered for
SIS energy HICs, where the $\Delta$(1232) is dominating,
denoted as $\Delta$ in short. For simplicity, the
isospin-independent part of the EoS, $e_0(u)$, for all other baryons
is taken to be the same as for the nucleons. The symmetry potentials for the
resonances are obtained through the constants of isospin coupling
(the Clebsch-Gordan coefficients) in the process of $\Delta$ [or
$N^*(1440)$] $\leftrightarrow \pi N$. For hyperons, based on the
analysis of the Lane potential \cite{Lan69,Dab99}, we simply take the
symmetry potential of hyperons to be nucleon-like.
However, the symmetry potential of excited states of hyperons
is not considered, for lack of information.

Combining both, resonances and hyperons, we express the
symmetry potential in a unified form, which reads as
\begin{equation}
v_{\rm sym}^{\rm B}=\alpha\, v_{\rm sym}^n+\beta\, v_{\rm sym}^{\rm p} ,\label{vx}
\end{equation}
where the values of $\alpha$ and $\beta$ for different baryons (B)
are listed in Table \ref{tab1}. From this table we can see that
the symmetry potentials of $\Delta^-$ and $\Sigma^{-}$ are
neutron-like and those of $\Delta^{++}$ and $\Sigma^{+}$ are
proton-like. On the other hand, the symmetry potentials of
$\Delta^0$, $\Delta^+$, $N^*$(1440), $\Sigma^0$, and $\Lambda$ are
a mixture of the neutron and proton symmetry potentials.

\begin{table}
\caption{Values $\alpha$ and $\beta$ for the symmetry potentials (Eq.\ \ref{vx}) of different nucleon and $\Delta$ resonances, as well as hyperons.}

\begin{tabular}{|l|cc||l|cc|}
\hline\hline B  & $\alpha$ & $\beta$ & B  & $\alpha$ & $\beta$  \\
\hline
{N$^*$}$^0$(1440)&$1/3$ & $2/3$&$\Lambda$ & $1/2$ & $1/2$\\
{N$^*$}$^+$(1440)&$2/3$ &$1/3$&$\Delta^-$&$1$ & $0$\\
$\Sigma^-$ & $1$ & $0$& $\Delta^0$ &$2/3$ & $1/3$\\
$\Sigma^0$ & $1/2$ & $1/2$& $\Delta^+$&$1/3$ & $2/3$ \\
$\Sigma^+$ & $0$ & $1$&$\Delta^{++}$ &$0$ & $1$\\
\hline\hline

\end{tabular}
\label{tab1}
\end{table}

In the standard UrQMD model (version 1.3), the cascade mode is usually adopted for produced mesons. In order to investigate the phase-space distributions of
pions, especially the transverse momentum distributions in intermediate energy HICs, the Coulomb potential of mesons
should be considered explicitly \cite{Pel97,LiB05,Gai04}.
Thus, in this paper, we also investigate the contribution of the
Coulomb potentials between mesons and baryons. Other
mean-field potentials of mesons are not yet considered.

\section{Effects of the symmetry potential on the dynamics of nucleons and $\Delta$'s}
\begin{figure}
\includegraphics[angle=0,width=0.8\textwidth]{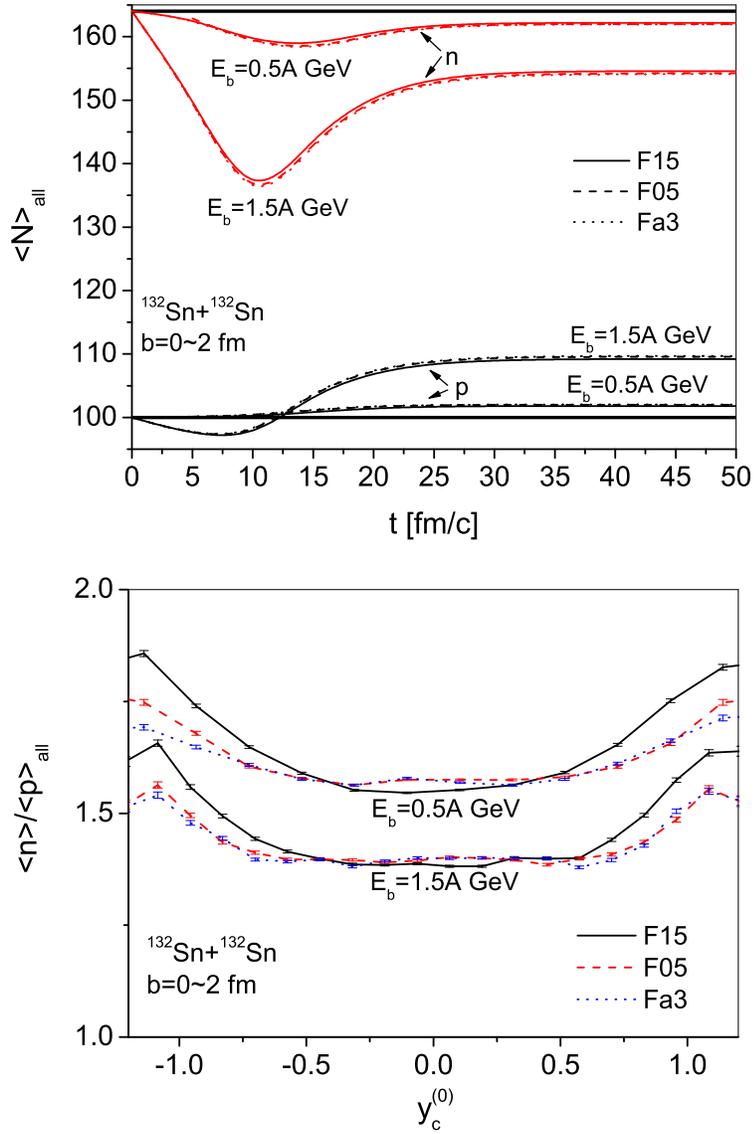}
\caption{Top: Time evolution of the neutron and proton
numbers for central $^{132}$Sn+$^{132}$Sn collisions at beam
energies $0.5A$ and $1.5A$ GeV. For the "hard" EoS, three
different forms of the symmetry potential are used (see text). The
two horizontal solid (thick) lines represent the initial neutron
and proton numbers of the system. Bottom: normalized
rapidity distribution of the neutron/proton ratio of all
nucleons at freeze-out time (see text). The error bars show
the statistical error.} \label{fig2}
\end{figure}

First of all, let us discuss the basic consequences of the density
dependence of the symmetry potentials in neutron-rich HICs at SIS
energies. Fig.\ \ref{fig2} (top) shows the time evolution of
the neutron and proton numbers for the symmetry potentials
F15, F05, and Fa3. Here, we do not distinguish whether the nucleons are free or bound in heavier fragments.
The beam energies are $E_{\rm b}=0.5A$ and $1.5A$ GeV and central collisions ($b=0-2\ {\rm fm}$) are chosen. From this plot, we notice that independent of the choice of the density dependence of the symmetry
potential, the neutron numbers reach their minima at $\sim\ 15\ {\rm fm}/c$ and
$10\ {\rm fm}/c$, for the beam energies
$E_{\rm b}=0.5A$ and $1.5A$ GeV, respectively. At this time the central nuclear
density reaches its maximum and some of the neutrons are excited
(to N$^*$ or $\Delta$) or converted to other baryons (e.g., hyperons) through various baryon-baryon or meson-baryon
collisions. We also notice that the higher the beam
energy, the more excited baryon states are produced. There are obviously less excited protons than neutrons at the compression stage. In other words, at
the compression stage, the effect on the neutron number is stronger
than on the proton number, which is due to the neutron-rich
nuclear environment. After the decay of these unstable baryons,
the emitted mesons (especially, the pions) alter the
neutron/proton ratio (the neutron number
decreases, and, correspondingly, the proton number increases), and
hence the ratio tends to become unity with an increase of the
beam energy. In addition the effect of the
density-dependent symmetry potentials on the dynamical process is smaller at higher
energies \cite{LiQ05}. With a softer symmetry potential Fa3, more neutrons are
excited at the compression stage, which is clearly due to more
neutrons being kept by the softer symmetry potential in the high
density region, as implied from Fig.\,\ref{fig1}.

\begin{figure}
\includegraphics[angle=0,width=0.8\textwidth]{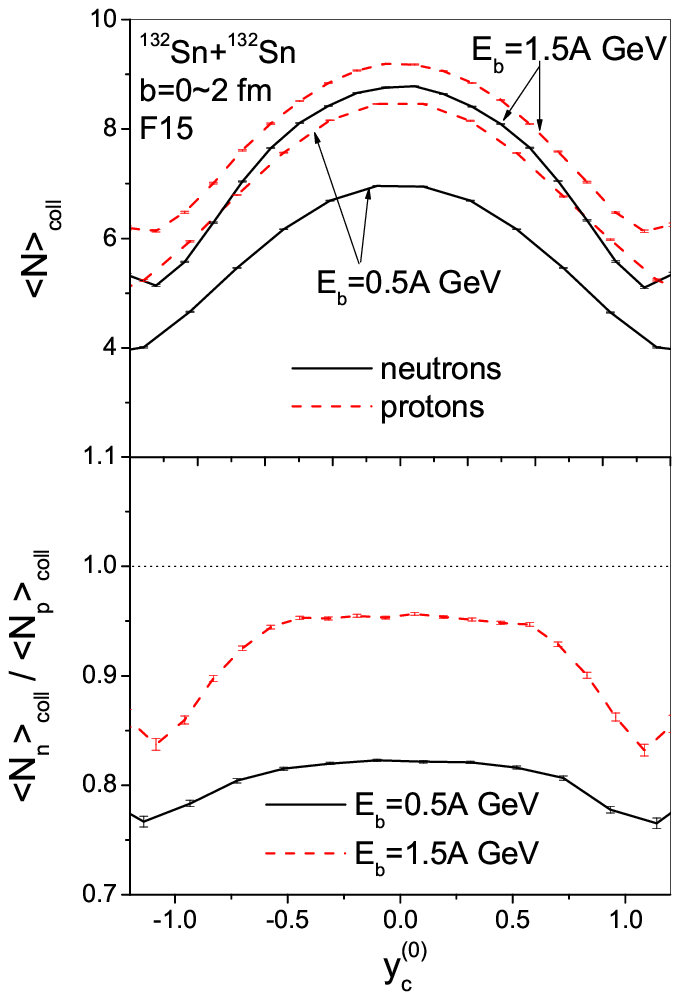}
\caption{Top: Normalized rapidity distributions of the
collision number of neutrons and protons at energies $E_{\rm b}=0.5A$
and $1.5A$ GeV. The case of the symmetry potential F15 is
shown. Bottom: The corresponding ratios between the
collision numbers of neutrons and protons.} \label{fig3}
\end{figure}

In Fig.\ \ref{fig2} (bottom), the
neutron/proton ratio of the all (free and bound) nucleons at freeze-out
time ($t_{\rm f}=50\ {\rm fm}/c$, the maximum time used in the top
plot) is shown as a function of the normalized rapidity
($y_c^{(0)}=y_c/y_{\rm beam}$, where $y_c$ is the rapidity of the particle in the center-of-mass
system and $y_{\rm beam}$ is the beam rapidity).
At $E_{\rm b}=0.5A$ GeV, the density
dependence of the symmetry potential influences the neutron/proton
ratio strongly, whereas at $E_{\rm b}=1.5A$ GeV the sensitivity is reduced, especially at midrapidity. This happens not only
due to the relatively weak effect of the EoS, as compared to the
two-body collision dynamics, but also due to the nucleon-nucleon cross sections at
higher energies that depend less on isospin. Secondly, in the projectile and target rapidity
regions, the $n/p$ ratio is more sensitive to the density dependence of the symmetry
potential as compared to the midrapidity region. Less collisions take
place in the projectile and target rapidity regions than in the
midrapidity region (see also, Fig.\ \ref{fig3}). Thirdly, the $n/p$ ratio is larger with a softer symmetry potential Fa3 than that with F15, and vice versa in the
projectile-target regions. The
nucleons in the midrapidity region represent mainly the behavior of symmetry potential at high
densities, where more neutrons
are kept with a soft symmetry potential. We should also notice that at midrapidity, the difference of the ratios with the
F05 and Fa3 symmetry potentials is almost negligible. From Fig.\ \ref{fig1} we can see that F05 is always stiffer than Fa3 at both low and high densities, thus the $n/p$ ratio is also influenced by the density dependence of the symmetry potential at subnormal densities. Therefore, in order to investigate any isospin
dependences, it is very important to know the
correct density-dependent form of the symmetry potential at {\it both}, low and high densities.

Fig.\ \ref{fig3} shows the calculated rapidity distribution of the
collision number of neutrons and protons (upper plot), and the
corresponding ratios (lower plot). Evidently, at both
energies, the collision number of nucleons has a strong maximum at midrapidity. The collision number increases with beam energy,
especially for neutrons because of the neutron-rich system. The average proton collision
number is always larger than the average neutron collision number
because of the differences in the
nucleon-nucleon cross sections at intermediate energies \cite{PDG96}.
When the beam energy increases from $0.5A$ to $1.5A$ GeV, this
isospin effect of the nucleon-nucleon cross section
is largely reduced, so that the ratio between the neutron and
proton collision numbers approach unity, which is shown in the
lower part of Fig.\ \ref{fig3}. In the midrapidity region, the
collision ratio of neutrons and protons is $\sim 82\%$ at
$E_{\rm b}=0.5A$ GeV, which increases to $\sim 95\%$ at $E_{\rm b}=1.5A$ GeV.

\begin{figure}
\includegraphics[angle=0,width=0.8\textwidth]{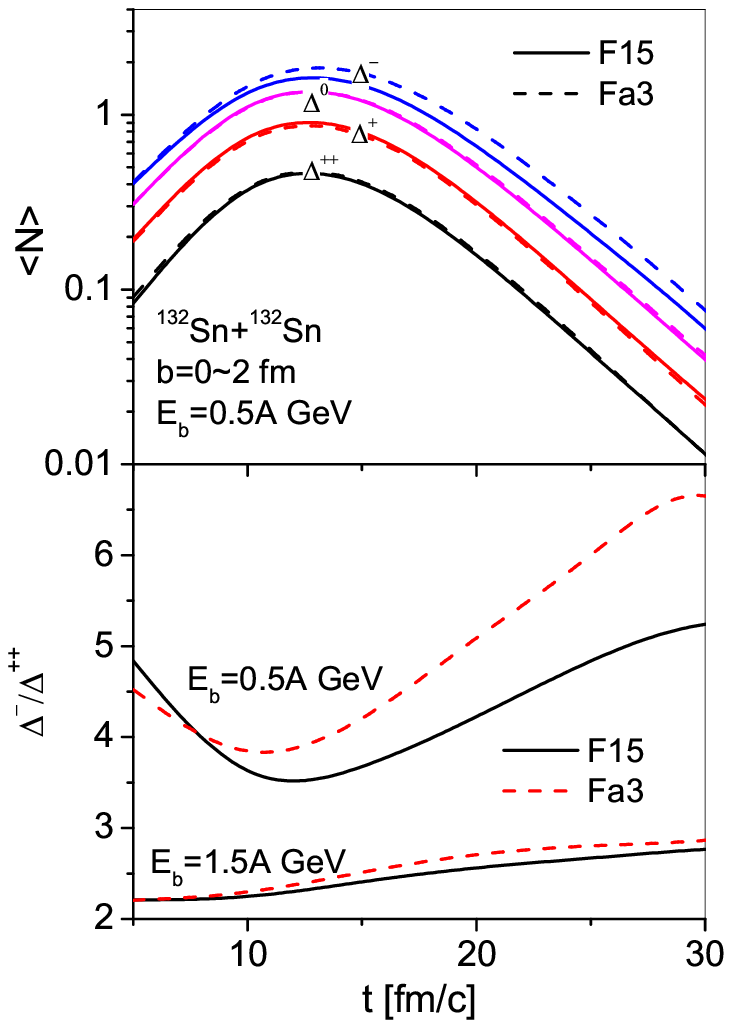}
\caption{Top: Time evolution of $\Delta$ abundancies, for the symmetry potentials F15 and Fa3 at
the beam energy $E_{\rm b}=0.5A$ GeV. Bottom: Time evolution of the $\Delta^-/\Delta^{++}$ ratio for the two symmetry potentials and
at two beam energies $E_{\rm b}=0.5A$ and $1.5A$ GeV.} \label{fig4}
\end{figure}

Since, in the SIS energy region, $\pi$-mesons are mainly produced
from decaying $\Delta$'s, it is also important to investigate
the dynamics of the $\Delta$'s with respect to the symmetry
potential. Fig.\ \ref{fig4} shows the time evolution of the
various components of $\Delta$ (upper plot) and the ratio
$\Delta^-/\Delta^{++}$ (lower plot) for the symmetry potentials
F15 and Fa3 and for central $^{132}$Sn+$^{132}$Sn collisions at
the two energies $E_{\rm b}=0.5A$ and $1.5A$ GeV. In the upper
plot, only the case $E_{\rm b}=0.5A$ GeV is shown since, at
$E_{\rm b}=1.5A$ GeV, the effect of the density dependence of the
symmetry potentials on the $\Delta$ production is almost
negligible. This is supported by the time evolution of the
ratio $\Delta^-/\Delta^{++}$ at $E_{\rm b}=1.5A$ GeV in the lower plot.

From the upper part of Fig.\ \ref{fig4}, it is clear that the
effect of the density-dependent symmetry potentials on the time
evolution is strongest for the $\Delta^-$ production, like for
neutrons in Fig.\,\ref{fig2}. From the time evolution of the
$\Delta^-/\Delta^{++}$ ratio in the lower plot of Fig.\ \ref{fig4}, we
further notice that, similar to the case of nucleons, the effect
of the density-dependent symmetry potentials on the
$\Delta^-/\Delta^{++}$ ratio is largely reduced at the higher beam energy $1.5A$ GeV. And the time evolution of
this ratio is nearly flat since, at this energy, the effect of the
Coulomb and the symmetry potentials becomes small. At the lower beam energy $0.5A$ GeV and during the compression stage ($t\lesssim 12\
{\rm fm}/c$), the $\Delta^-/\Delta^{++}$ ratio decreases
with time while it increases at the later stage, the expansion stage. This is apparently due to the dynamically
mutual interaction between the Coulomb and the
symmetry potentials: along the time evolution at the compression stage, more neutrons are pushed out because the relative strength of the symmetry
potential as compared to the Coulomb
potential increases. This can be understood from the study on the ratio of preequilibrium neutron number to proton number in the intermediate-energy neutron-rich HICs in Ref.\ \cite{Liu:2001ud}: this ratio is larger than the initial neutron/proton ratio of the colliding system; at the expansion stage, more protons
are pushed out due to the stronger Coulomb potential as compared to the symmetry potential. Note
that the increase of the $\Delta^-/\Delta^{++}$ ratio at the late stage is stronger for the symmetry potential Fa3 than for F15. The time evolution of the $\Delta$
abundancies, depending on the symmetry potential, should also affect
the production of pions.

\begin{figure}
\includegraphics[angle=0,width=0.8\textwidth]{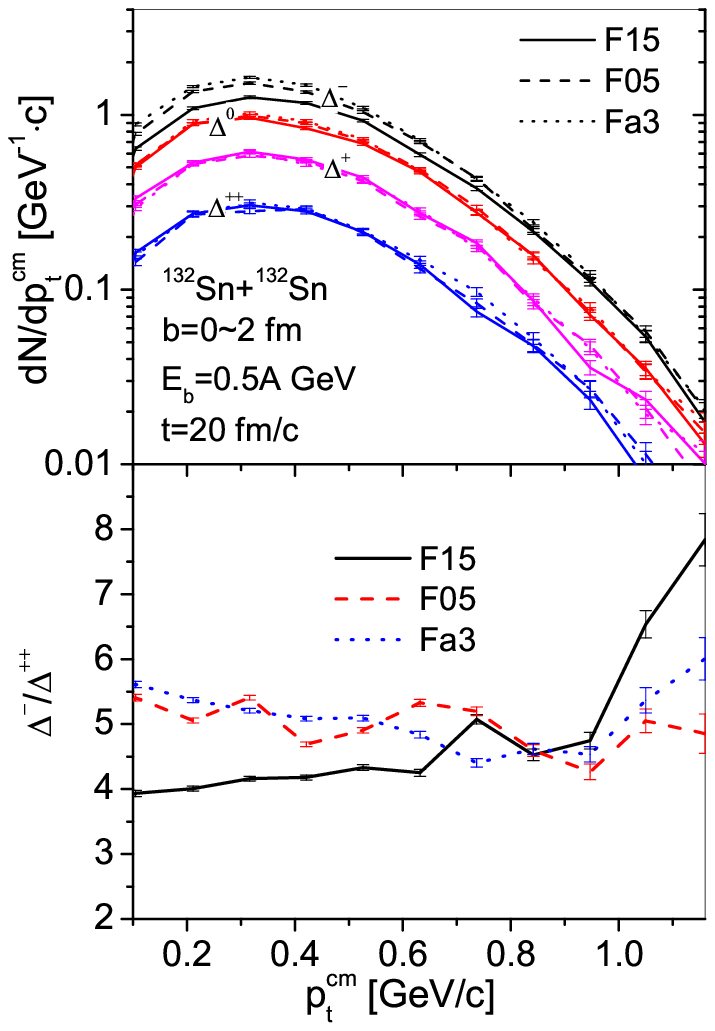}
\caption{Top: Transverse momentum distributions of various $\Delta$ components for the symmetry potentials F15,
F05, and Fa3, at $E_{\rm b}=0.5A$ GeV, and $t=20\ {\rm fm}/c$. Bottom: The
corresponding $\Delta^-/\Delta^{++}$ ratios.}
\label{fig5}
\end{figure}

Besides the $\pi$-yields, the pion spectra should also be influenced by the decay of $\Delta$'s. Fig.\,\ref{fig5} shows the transverse momentum spectrum of
$\Delta$'s (upper plot) and the $\Delta^-/\Delta^{++}$ ratios
(lower plot) at $t=20\ {\rm fm}/c$ and $E_{\rm b}=0.5A\ {\rm GeV}$, for the symmetry
potentials F15, F05, and Fa3. We notice from the upper part of
Fig.\ \ref{fig5} that, similar to the neutrons in
Fig.\ \ref{fig2}, the effect of different density-dependent symmetry
potentials is strongest for the $\Delta^-$. From
the lower plot of Fig.\ \ref{fig5} we can see that, at lower $p_t^{\rm cm}$
($p_t^{\rm cm}\lesssim 0.8\ {\rm GeV}/c$), the $\Delta^-/\Delta^{++}$ ratios
are quite different for the symmetry potentials F15 and F05 (or
Fa3), and the results of the F05 and Fa3 symmetry potentials are nearly
the same. These results are similar to the rapidity distribution of the
neutron/proton ratio, shown in Fig.\ \ref{fig2}.
$\Delta$'s with low transverse momenta are not produced from very high densities, namely around normal density, where
the difference between F05 and Fa3 is quite small and the effect of F05 and Fa3 on the $\Delta^-/\Delta^{++}$ ratio are cancelled strongly between low and high densities. For  $p_t^{\rm cm} \gtrsim 0.8\ {\rm GeV}/c$, the differences between the results
of F05 and Fa3 become distinguishable because $\Delta$'s with the high
transverse momenta are mainly produced from the central
high-density region. Thus the
$\Delta^-/\Delta^{++}$ ratio at high transverse momenta can
provide clearer information on the density dependence of the
symmetry potential. In the next section, we investigate the question to what extent this property could be transferred to $\pi$'s, especially when the
contribution of the Coulomb potential of mesons is also
considered.

\section{Effects of the symmetry potential on the $\pi$ and $K$ production}

\begin{figure}
\includegraphics[angle=0,width=0.8\textwidth]{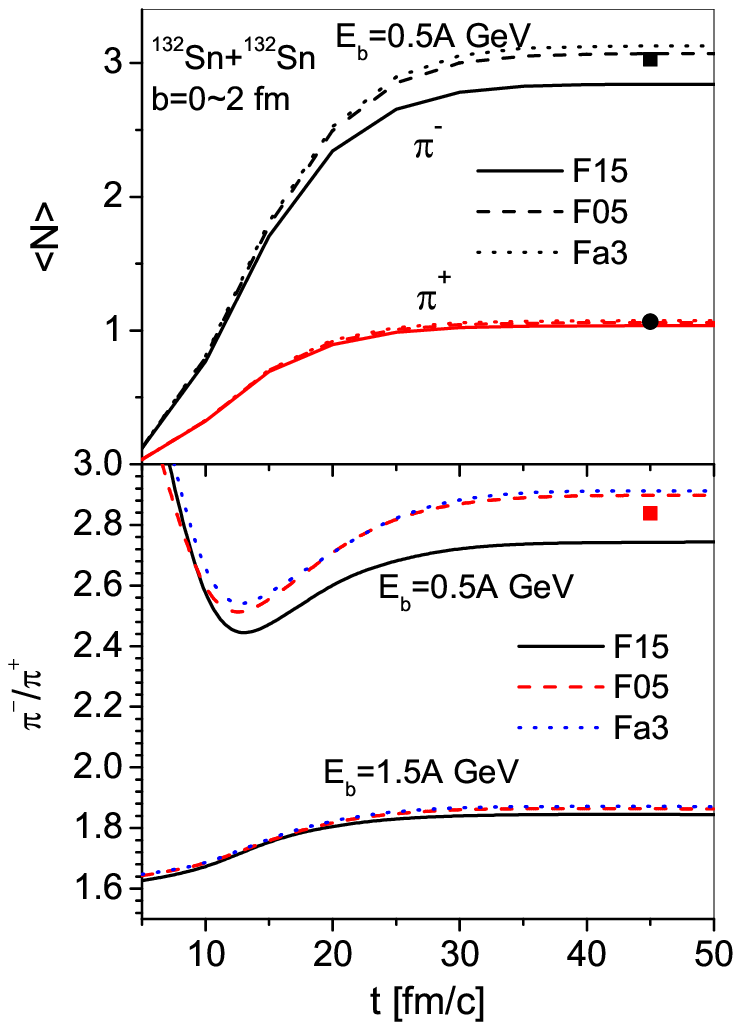}
\caption{Time evolution of the $\pi$ abundancies (top) and
the $\pi^-/\pi^+$ ratios (bottom) for the symmetry potentials
F15, F05, and Fa3, at the beam energy $E_{\rm b}=0.5A$ GeV. The Coulomb potential of mesons is switched off, except for the solid dots
for the case F05 at $E_{\rm b}=0.5A$ GeV. In the lower plot, the
$\pi^-/\pi^+$ ratio at $E_{\rm b}=1.5A$ GeV is also shown.} \label{fig6}
\end{figure}

Fig.\ \ref{fig6} shows the time evolution of the $\pi$ abundancies
(upper plot) and their ratios (lower plot) for the
density-dependent symmetry potentials F15, F05, and Fa3. Here, the
Coulomb potential of mesons is switched off. However, for
comparison, we also provide the results (for the case of F05 and at
$E_{\rm b}=0.5A\ {\rm GeV}$) when the Coulomb potential of mesons is taken into account (solid points in Fig.\ \ref{fig6} at time $t=45\ {\rm fm}/c$). Apparently, the effect of
the Coulomb potential of mesons on the $\pi$-yields is quite small. We also find that the effect of the different symmetry potentials on the $\pi^-/\pi^+$ ratio is hardly affected by the Coulomb potential of mesons although it alters the absolute value of the ratio.
The number of $\pi^-$ grows much stronger than the number of $\pi^+$, almost independent of the choice of the symmetry
potentials F05 and Fa3. The $\pi^-$-yields are, however, smaller
for F15. So is the $\pi^-/\pi^+$ ratio at
$E_{\rm b}=0.5A\ {\rm GeV}$. On the other hand, for the higher beam energy
$E_{\rm b}=1.5A$ GeV, the ratios are almost independent of the
symmetry potentials. (These results are clearly
due to the properties of $\Delta$'s, shown in Figs.\ \ref{fig4} and
\ref{fig5}). Therefore, it is necessary to pay attention to the energy dependence of the isospin effect on the $\pi^-/\pi^+$ ratios as well.

\begin{figure}
\includegraphics[angle=0,width=0.8\textwidth]{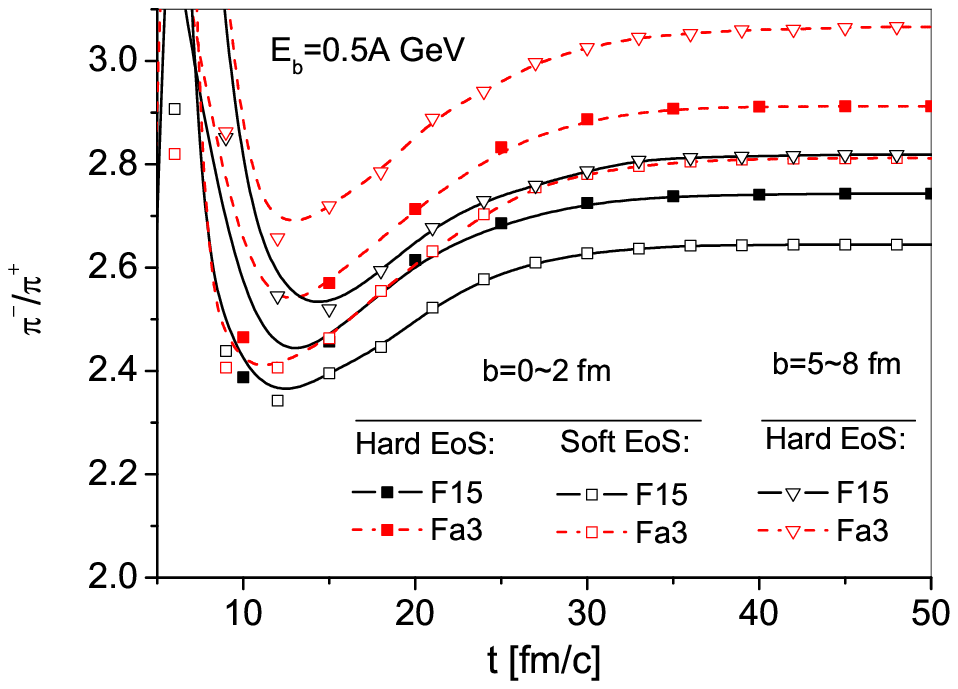}
\caption{$\pi^-/\pi^+$ ratios from $^{132}$Sn+$^{132}$Sn
reactions at $0.5A$ GeV, using a 'hard' and a 'soft' EoS
with impact parameters $b=0 - 2\ {\rm fm}$ and $5 - 8\ {\rm fm}$. The
symmetry potentials used are F15 and Fa3.} \label{fig7}
\end{figure}

Next, we show the dependence of the $\pi^-/\pi^+$ ratio on various
isospin-independent EoS and impact parameter $b$. In view of the
above noted beam-energy dependence, we only consider
the lower beam energy $E_{\rm b}=0.5A$ GeV in the following. Fig.\ \ref{fig7}
shows the results for two EoS and two centralities. The effect of the density-dependent symmetry potentials on
the $\pi^-/\pi^+$ ratios is almost not affected by the uncertainty
of the isospin-independent EoS. That is, the absolute shift in the $\pi^-/\pi^+$ ratio due to different symmetry potentials remains almost the same, independent of the stiffness of the isospin-independent EoS. On the other hand, at larger impact parameters ($b=5\sim 8\ {\rm fm}$),
the $\pi^-/\pi^+$ ratio is even more sensitive to the density dependence of the symmetry potential.

\begin{figure}
\includegraphics[angle=0,width=0.8\textwidth]{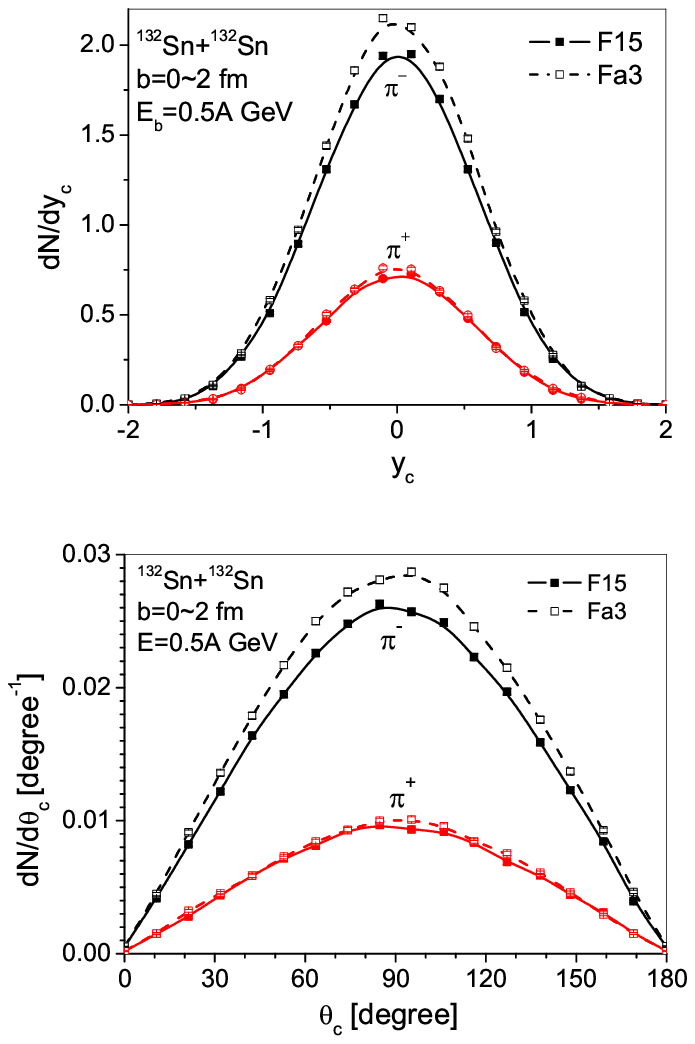}
\caption{Rapidity (top) and the polar angle (bottom)
distributions of $\pi^-$ and $\pi^+$ in central $^{132}$Sn+$^{132}$Sn collisions at $E_{\rm b}=0.5A\ {\rm GeV}$, for the symmetry
potentials F15 and Fa3.}
\label{fig8}
\end{figure}

The rapidity $y_c$ and the polar angle $\theta_c$ (in the center-of-mass system) distributions of $\pi^-$
and $\pi^+$ for various density-dependent symmetry potentials are
shown in Fig.\ \ref{fig8}. The
influence of the density-dependent symmetry potentials on the
$\pi$-multiplicity distributions is strongest at midrapidity and at
polar angles around $90^0$. This holds in particular for the
$\pi^-$ distribution, which was also seen in \cite{LiB05} for the
kinetic energy spectrum of pions in central
$^{132}$Sn+$^{124}$Sn reactions at $E_{\rm b}=0.4A\ {\rm
GeV}$.

\begin{figure}
\includegraphics[angle=0,width=0.8\textwidth]{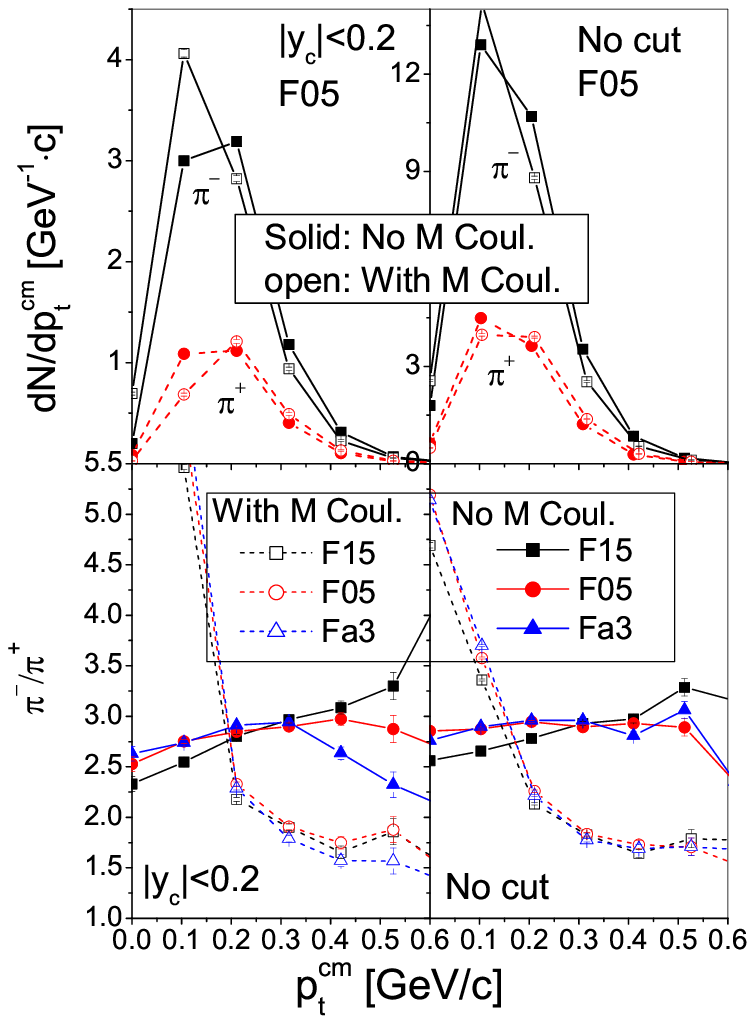}
\caption{Upper two plots: Transverse momentum distribution of
$\pi^-$ and $\pi^+$ mesons with (left) and without (right) rapidity cut $|y_c|<0.2$. The results with and
without Coulomb potential of mesons (denoted as "M Coul.") are
shown. The symmetry potential used is F05. Lower two plots: The corresponding $\pi^-/\pi^+$ ratios, but for three different
density-dependent symmetry potentials. } \label{fig9}
\end{figure}

The upper two plots of Fig.\ \ref{fig9} show the transverse momentum
distributions of $\pi^-$ and
$\pi^+$, with and without the contribution of the Coulomb
potentials of mesons, as well as with a rapidity-cut
($|y_c|<0.2$) (left plots) and without any kinematical cut (right plots), for the symmetry
potential F05. This is important because, in recent years,
several experiments on charged pion production at intermediate
energies were performed by the FOPI Collaboration at GSI
where the possible contribution of the Coulomb
potentials of pions to the $\pi$-transverse momentum distribution
was also discussed \cite{Pel97,Hon05}. Although, as shown in Fig.\ \ref{fig6}, the effect of the Coulomb potential of mesons on the
total $\pi$ yield is small, Fig.\ \ref{fig9} demonstrates that it strongly
influences the momentum distribution; the Coulomb potential of mesons always shifts the
$\pi^-$ to lower, and the $\pi^+$ to higher $p_t^{\rm cm}$ for the
neutron-rich systems, in line with Ref.\,\cite{LiB05}. With the
rapidity-cut $|y_c|<0.2$, this effect becomes even more pronounced;
as a result, the transverse momentum distribution of the
$\pi^-/\pi^+$ ratio at low $p_t^{\rm cm}$ becomes steeper than the
one without the rapidity cut (lower plots in Fig.\ \ref{fig9} show the corresponding $\pi^-/\pi^+$ ratios for the
symmetry potentials F15, F05, and Fa3). If the Coulomb potential of mesons is switched off, the $\pi^-/\pi^+$ ratio without rapidity-cut
is nearly constant at low transverse momenta
($p_t^{\rm cm}<0.3\ {\rm GeV}/c$). With the rapidity-cut ($|y_c|<0.2$) it
increases weakly with $p_t^{\rm cm}$ because of the contribution of the
Coulomb potential of $\Delta$'s. At higher transverse momenta
($p_t>0.4\ {\rm GeV}/c$), and with rapidity-cut (lower-left
plot of Fig.\ \ref{fig9}), the effect of the density-dependent
symmetry potentials on the $\pi^-/\pi^+$ ratios is noticeable; the
difference is also obvious for F05 and Fa3. It is recalled that a similar
phenomenon can be seen in Fig.\ \ref{fig5} for the
$\Delta^-/\Delta^{++}$ ratios. However, when the Coulomb
potentials of mesons are taken into account, the effect of the
density-dependent symmetry potentials on the $\pi^-/\pi^+$ ratios
at high transverse momenta is largely reduced, regardless of a rapidity cut. It was pointed out in \cite{LiB05}
that the Coulomb potential is stronger than the symmetry
potential, and the Coulomb potential acts directly on the charged
pions but the symmetry potential does not, such that the $\pi$'s
emitted from the dense region, which have higher $p_t^{\rm cm}$, may
be influenced much more by their Coulomb potential than by the
symmetry potential. Considering the limit of experiments so far,
we claim that the $\pi^-/\pi^+$ production ratio at $p_t^{\rm cm}=0.1\ {\rm GeV}/c$ is a suitable candidate for probing the density
dependence of the symmetry potential, which is discussed in
more detail in \cite{LiQ052}.

\begin{figure}
\includegraphics[angle=0,width=0.8\textwidth]{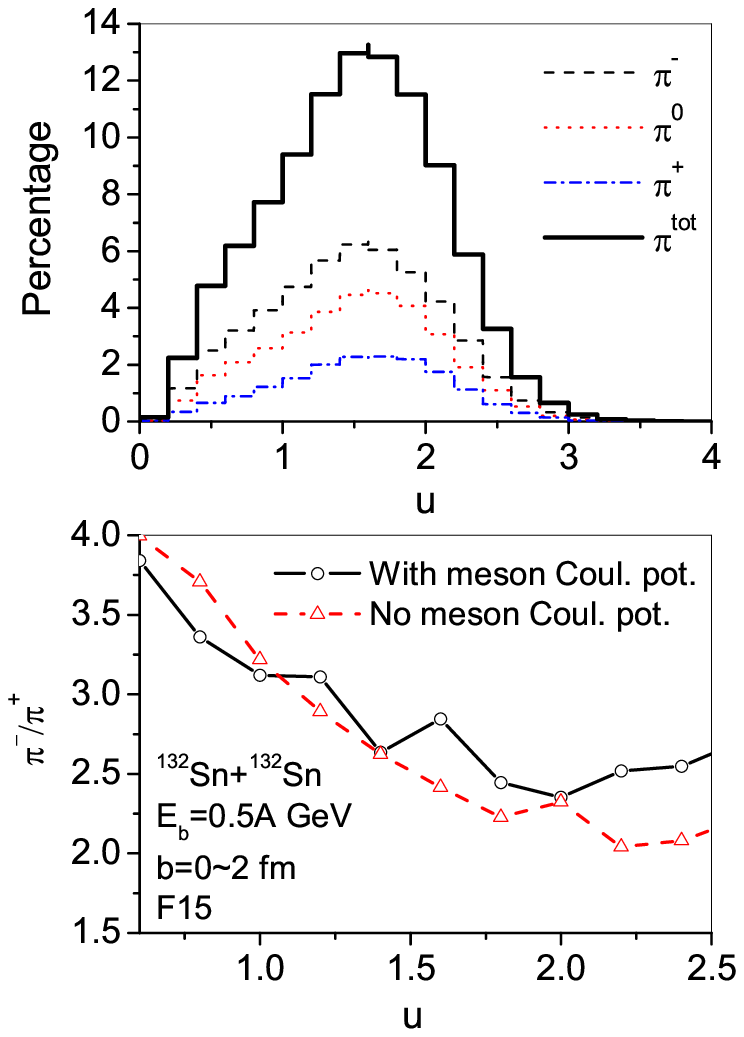}
\caption{Top: Percentages of the $\pi$'s as a function of
reduced density, with the Coulomb potential of mesons included. Bottom: the corresponding density distribution of the $\pi^-/\pi^+$
ratio of the percentages, with or without the Coulomb potential
of mesons.} \label{fig10}
\end{figure}

Fig.\ \ref{fig10} shows the density distribution of pion emission, with and
without Coulomb potential of mesons. In the upper plot, the
percentages of each component of $\pi$ and of the total $\pi$ yield
are shown as a function of the reduced density $u$, in steps of
$\Delta u=0.2$, with the Coulomb potential of mesons included.
Most pions are emitted from densities higher
than normal nuclear density (especially from the density
region $u \sim 1-2$); a few pions ($\sim 16\%$) are emitted
from subnormal densities. This follows from Fig.\ \ref{fig5} where quite a few $\Delta$'s are transported close to normal nuclear density. As these $\Delta$'s decay, some of them
also appear at subnormal densities.

In the lower part of Fig.\ \ref{fig10}, the ratios of the
percentages of $\pi^-$ and $\pi^+$ are shown as a function of
the reduced density $u$, with and without the Coulomb potential of mesons.
The ratios are always larger at lower densities
than at higher densities since at the late stage, where the nuclear density decreases, the $\Delta^-/\Delta^{++}$ ratio (shown in Fig.\ \ref{fig4}) increases with time. When the Coulomb potential of mesons
is taken into account, more $\pi^-$ mesons are decelerated because of
the larger amount of the negatively charged particles in the low
density region, and are thus emitted with low momenta which
leads to the rise of the $\pi^-/\pi^+$ ratios at low $p_t^{\rm cm}$,
as shown in Fig.\ \ref{fig9}.

\begin{figure}
\includegraphics[angle=0,width=0.8\textwidth]{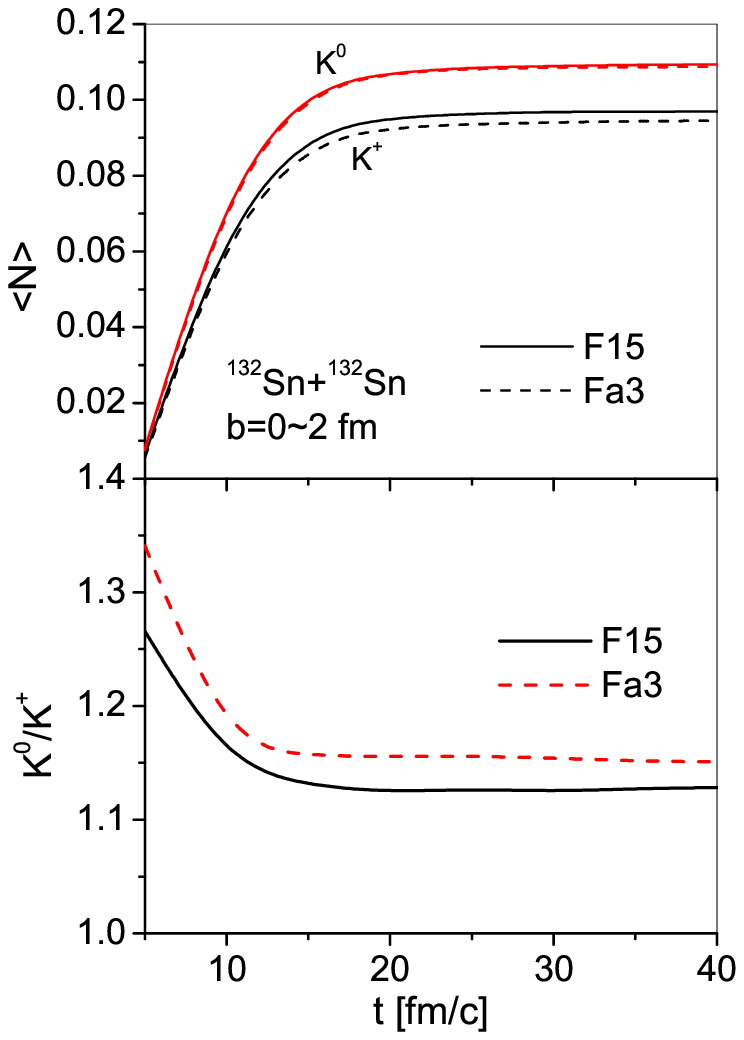}
\caption{Top: Time evolution of the K$^0$ and K$^+$ abundancies
for central $^{132}Sn+^{132}Sn$ collisions at a beam energy
$1.5A$ GeV and with the symmetry potentials F15 and Fa3. Bottom: the corresponding time evolution of the $K^0/K^+$ ratios.}
\label{fig11}
\end{figure}

Besides the $\pi^-/\pi^+$ ratios, the $K^0/K^+$ ratios
are also thought to be a suitable candidate to probe the
density dependence of the symmetry potential in dense nuclear
matter, as already discussed in the Introduction. Fig.
\ref{fig11} shows the time evolution of the $K^0$ and $K^+$
abundancies, as well as their ratios $K^0/K^+$, for central
$^{132}Sn+^{132}Sn$ collisions at beam energy $1.5A$ GeV and for
the symmetry potentials F15 and Fa3. Similar to the
$\pi$ production, the effect of different density
dependences of the symmetry potential on the kaon yields is quite
reduced at higher energies.

In SIS energy HICs, kaons are emitted mainly from the high
density region and at the early stage of the reaction. The
dominant production channels are  $BB\rightarrow BB^*\rightarrow BKY$ and
$B\pi\rightarrow B^* \rightarrow KY$ (here $Y$ represents a
hyperon). With a softer symmetry potential, in the neutron-rich
nuclear medium, more neutrons (protons) are shifted to the high (low)
density region and hence more negatively (less positively) charged
$B^*$ are produced, and the $K^0/K^+$ ratio increases. Thinking of the energy dependence,
the kaon production at energies much lower than its threshold ($\sim 1.58$ GeV from the nucleon-nucleon interaction in free space) might be more sensitive to the density
dependence of the symmetry potential. As an example, we have also
calculated the kaon yields from the reaction $^{208}Pb+^{208}Pb$
at $E_{\rm b}=0.8A$ GeV and $b=7\sim 9\ {\rm fm}$ with the symmetry potentials
F15 and Fa3: the $K^0/K^+$ ratio for F15 is about $1.25$, whereas it is about $1.4$
for Fa3. We should point out that these results on kaons might be relatively rough since we do not consider any kaon-nucleon mean-field potential \cite{LiG95,Fuchs02,Nek02}. As far as we know, there is no explicit calculation about the difference of $K^+$- and $K^0$-nucleon potentials in the nuclear medium.

\section{Summary}

In this paper, we have investigated the role of the
density-dependent symmetry potential in heavy ion collisions
(HICs) at SIS energies, based on the UrQMD model (v1.3).
The contribution of the Coulomb potential of mesons has been studied, in order to better understand the effects of the
density-dependent symmetry potential on the dynamics of pions
produced in neutron-rich HICs. The calculated results show a strong beam-energy dependence of the effect of the
density-dependent symmetry potentials on the production ratios
$\Delta^-/\Delta^{++}$ as well as $\pi^-/\pi^+$. The uncertainty
in the isospin-independent EoS alters the values of both the pion
yields and the $\pi^-/\pi^+$ ratios, but has almost no effect on
the role of the density-dependent symmetry potential on the
$\pi^-/\pi^+$ ratios. The impact parameter is found to be important
for the $\pi^-/\pi^+$ ratios; for a larger impact parameter
the effect of different density-dependent symmetry potentials on the
$\pi^-/\pi^+$ ratios becomes stronger.

The Coulomb potential of mesons changes the transverse momentum
distribution of the $\pi^-/\pi^+$ ratios significantly, though it
leaves the total $\pi^-$ and $\pi^+$ yields almost unchanged. We
find that the negatively charged pion yields, especially at
midrapidity and low transverse momenta, as well as the
$\pi^-/\pi^+$ ratios at low transverse momenta, could be sensitive
to the density-dependent symmetry potential in a dense nuclear
matter. Quite a few pions are still produced at subnormal
densities and thus are affected by the density-dependent symmetry
potential at subnormal densities. These studies are of interest
for forthcoming experiments at RIA (USA) and FAIR/GSI
(Germany).

In this work, we have also investigated the
yields of K$^0$ and K$^+$ mesons and the ratios $K^0/K^+$
from neutron-rich HICs at $E_{\rm b}=1.5A$ GeV. It is shown that
they do not seem to be suitable to investigate the density dependence of the
symmetry potential in dense nuclear matter. However, kaon
production at energies much lower than threshold might improve the situation and is
worth investigating further. Meanwhile, it is necessary to consider the difference of the $K^+$- and $K^0$-nucleon potentials in the nuclear medium.

\section*{Acknowledgments}
We would like to acknowledge valuable discussions with S. Schramm and A. Mishra. Q. Li thanks the Alexander von Humboldt-Stiftung for a fellowship. RK Gupta
thanks Deutsche Forschungsgemeinschaft (DFG) for a Mercator Guest Professorship.
This work is partly supported by the National Natural Science
Foundation of China under Grant No.\ 10255030, by GSI, BMBF, DFG, and Volkswagen Stiftung.

\end{document}